\documentclass[aps,prl,twocolumn,superscriptaddress,floatfix, showpacs]{revtex4}

\usepackage{graphicx}
\usepackage{amsmath}
\usepackage{amssymb}

\begin{document}

\title{Spin tunneling in junctions with disordered ferromagnets}

\author{P.~V.~Paluskar}
\email[]{p.v.paluskar@tue.nl}
\affiliation{Department of Applied Physics, Center for NanoMaterials and COBRA Research Institute, Eindhoven University of Technology,
P.O.~Box~513, 5600 MB Eindhoven, The Netherlands}
\author{J.~J.~Attema}
\author{G.~A.~de~Wijs}
\affiliation{Institute for Molecules and Materials, Faculty of Science, Radboud University Nijmegen, Toernooiveld 1, 6525 ED Nijmegen, The Netherlands}
\author{S.~Fiddy}
\affiliation{STFC Daresbury Laboratory, Warrington, Cheshire, WA4 4AD, UK}
\author{E.~Snoeck}
\affiliation{CEMES- CNRS,
29, rue Jeanne Marvig  B.P. 94347 F-31055 Toulouse Cedex 4, France}
\author{J.~T.~Kohlhepp}
\affiliation{Department of Applied Physics, Center for NanoMaterials
and COBRA Research Institute, Eindhoven University of Technology,
P.O.~Box~513, 5600 MB Eindhoven, The Netherlands}
\author{H.~J.~M.~Swagten}
\affiliation{Department of Applied Physics, Center for NanoMaterials
and COBRA Research Institute, Eindhoven University of Technology,
P.O.~Box~513, 5600 MB Eindhoven, The Netherlands}
\author{R.~A.~de~Groot}
\affiliation{Institute for Molecules and Materials, Faculty of Science, Radboud University Nijmegen, Toernooiveld 1, 6525 ED Nijmegen, The Netherlands}
\author{B.~Koopmans}
\affiliation{Department of Applied Physics, Center for NanoMaterials
and COBRA Research Institute, Eindhoven University of Technology,
P.O.~Box~513, 5600 MB Eindhoven, The Netherlands}

\date{\today}

\begin{abstract}
We provide compelling evidence to establish that, contrary to one's elementary guess, the tunneling spin polarization (TSP) of amorphous CoFeB is larger than that of highly textured fcc CoFeB. First principles atomic and electronic structure calculations reveal striking agreement between the measured TSP and the predicted s-electron spin polarization. Given the disordered structure of the ternary alloy, not only do these results strongly endorse our communal understanding of tunneling through AlO$_\textrm{x}$, but they also portray the key concepts that demand primary consideration in such complex systems.
\end{abstract}


\maketitle

Right from its inception, experimental and theoretical endeavors in electron tunneling have been dedicated to the understanding of the role of the electrode and barrier electronic structure. Not long after it was well-established that the density of states of a superconducting electrode was directly observable in tunneling through amorphous AlO$_\textrm{x}$ barriers~\cite{GiaeverPRL147}, tunneling spectroscopies to observe the influence of the electronic structure of semi-metallic electrodes were performed~\cite{Esaki}. For ferromagnetic films, one aspect of their electronic structure -- the tunneling spin polarization (TSP) -- was measured~\cite{Meservey}. Although some preliminary effort was undertaken to study the role of the band structure of ferromagnetic films in tunneling~\cite{RowellJAP}, no definitive observations were made till the advent of tunnel magnetoresistance (TMR) in magnetic tunnel junctions (MTJs). Then, Yuasa~\textit{et al.}~\cite{YuasaEPL} and LeClair~\textit{et al.}~\cite{LeClairPRLBS} experimentally demonstrated the influence of epitaxial Fe and textured Co films on TMR and tunneling conductance, respectively. The former established the change in TMR in Fe/AlO$_\textrm{x}$/Fe MTJs by growing Fe electrodes in different crystal orientations. The latter demonstrated the change in tunnel conductance of Co/AlO$_\textrm{x}$/Co MTJs at bias voltages where certain bands were known to exist in the electronic structure of fcc Co. Regarding the nature of the electronic wave functions that govern the tunneling probability through AlO$_\textrm{x}$, the dominance of the spherically symmetric s-like electrons has been experimentally demonstrated~\cite{YuasaScience,YuasaPRLCr}. Recently, spintronics has witnessed a rapid rise in the importance of amorphous ferromagnets like CoFeB. They have contributed to huge TMR in AlO$_\textrm{x}$~\cite{WeiJAP2007}~and MgO~\cite{YuasaAPL2005}~based MTJs. They have also been used to observe the novel spin-torque diode effect~\cite{TulapurkarNature} and facilitated record-low switching currents in spin-torque based MTJs~\cite{HayakawaJJAPSTT}. Although their emerging importance in spintronics is unquestionable, neither has there been a theoretical and experimental analysis of their atomic and electronic structure, nor has the impact of these properties on their TSP been investigated.

In this letter, we explore the correlation between ferromagnet morphology, its electronic structure and their combined impact on TSP. One unique aspect -- crystallization of amorphous CoFeB with a single high temperature anneal ($\gtrsim$ 250~$^\circ$C~\cite{PaluksarJPDAP,YuasaAPLCoFeB}) -- is exploited to study the structural, magnetic and TSP related properties of amorphous and crystalline CoFeB in the same sample. Indeed, such control on morphology is not accessible in elemental magnetic films. The high temperature anneal stipulates a crucial requirement for our junctions, viz. the barrier properties should not change after annealing to ensure comparison between the TSP of as-deposited and annealed CoFeB. Contrary to alternative barriers like MgO, AlO$_\textrm{x}$ barriers are known to exhibit no TSP related changes after anneals up to $T_a$=500~$^\circ$C~\cite{KantAPL,ParkinNatMat}. When the structure of Co$_{72}$Fe$_{20}$B$_8$ is intentionally transformed from amorphous to highly textured fcc, we notice that a correlated alteration of the CoFeB electronic structure is induced. Contrary to one's primary intuition, this alteration of the electronic structure manifests in an intrinsically larger TSP for amorphous CoFeB as compared to that of highly textured fcc~CoFeB. First principles atomic structure calculations of amorphous CoFeB are found to be consistent with extended x-ray absorption fine structure (EXAFS) measurements. Remarkably, electronic structure calculations based on this atomic structure exhibit a conspicuous agreement between the spin polarization (SP) of the s-electron density of states (DOS) and the experimentally measured TSP, both for amorphous and crystalline CoFeB. The calculations also reveal that the B sp-states get highly spin-polarized and make a significant contribution to the alloy SP. We would like to emphasize that such a quantitative agreement between theory and experiment for a complex amorphous/crystalline ternary alloy has not been reported before. Moreover, given the recent development in CoFeB based spintronic devices, first principles atomic and electronic structure calculations, especially those corroborating spin polarized tunneling experiments, have not been reported yet. Furthermore, these results endorse several earlier concepts, for example, the high sensitivity of the tunnel conductance to the ferromagnet-barrier interface~\cite{leclair01b}, and the dominance of s-electrons in tunneling through AlO$_\textrm{x}$ barriers~\cite{YuasaScience,YuasaPRLCr}.

\begin{figure}[!t]
    \vspace{-2 mm} \begin{center}
      \includegraphics[width=8.7 cm]{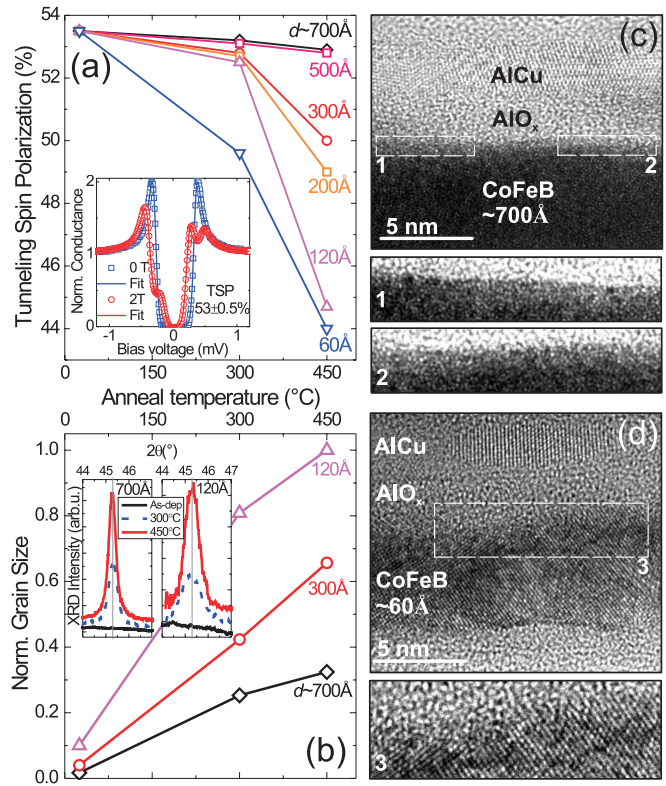}
      \caption{(color online) Inset in (a) shows a representative TSP measurement. The zero field curve ($\square$) shows the Al superconducting gap while the $2.0$ T ($\circ$) curve reveals the TSP of CoFeB when fit (solid lines) with Maki theory~\cite{maki64}.~(a)~TSP of CoFeB as a function of $T_a$ and \textit{d}.~(b)~The grain size perpendicular to the film plane is normalized to \textit{d} and plotted as a function of $T_a$. Insets show actual XRD data on as-deposited and annealed 700~\AA~and~120~\AA~films.~(c)~HRTEM micrograph of an Al/AlO$_\textrm{x}$/CoFeB (700~\AA)/Al junction after a 450~$^\circ$C anneal; see lower panels in (c) for magnified interface regions.~(d)~Similar junction, but with a 60~\AA~thick CoFeB.}\vspace{-6mm}
      \label{Fig1}
    \end{center}
\end{figure}

The inset in Figure~\ref{Fig1}a shows a representative TSP measurement for an as-deposited 120~\AA~CoFeB film using superconducting tunneling spectroscopy~\cite{Meservey}. Regardless of the CoFeB thickness (\textit{d}), for as-deposited samples, we consistently measure a TSP above 53~\%. However, as shown in Figure~\ref{Fig1}a, after annealing, the measured value of the TSP is strongly dependent on the thickness of the film and $T_a$. Evidently, thick films (700~\AA~and~500~\AA) show no significant change in the TSP after anneals above the crystallization temperature ($\gtrsim250^\circ$C). On the contrary, the TSP of progressively thinner films decreases systematically with the thickness of the films, especially for $T_a$=450$^\circ$C. One can rule out the formation of boron oxide at the barrier-ferromagnet interface or boron diffusion \textit{into} the tunnel barrier as a cause for this reduction in TSP since (a) both these processes are expected to contribute equally to the drop in TSP, \textit{regardless} of CoFeB thickness, (b) no significant change in junction resistance is observed, and (c) thermodynamically, AlO$_\textrm{x}$ is known to be a more stable oxide. Boron segregation \textit{away} from the interface can also be safely ruled out, as one might expect such a segregation to influence the TSP regardless of CoFeB thickness. These arguments also justify the use of low B content in this work. Moreover, the magnetic moment of CoFeB, independent of its thickness, does not show any significant post-anneal change. One would expect it to asymptotically proceed towards that of a comparable Co$_{80}$Fe$_{20}$ alloy, if boron would segregate.

A clue to the probable reason behind this change in the TSP of thin CoFeB films can be found in x-ray diffraction (XRD) measurements on films of corresponding thickness. In Figure~\ref{Fig1}b, the grain size perpendicular to the film plane, calculated using the Paul Scherrer formula, and normalized to the film thickness, is plotted as a function of $T_a$. This plot indicates that, in progressively thinner films, the grain sizes become comparable to the film thickness after the anneal. For $T_a$=450$^\circ$C and~\textit{d}=120~\AA, the average grain size is almost equal to the film thickness suggesting the presence of crystalline CoFeB at the interface with the AlO$_\textrm{x}$ barrier. This hypothesis is substantiated by high resolution transmission electron micrographs (HRTEM). Figure~\ref{Fig1}c shows a junction with a 700~\AA~CoFeB layer, while Figure~\ref{Fig1}d corresponds to a 60~\AA~CoFeB layer, both annealed at 450$^\circ$C. For the 700~\AA~film, a close inspection of the barrier-ferromagnet interface region shows hardly any crystalline CoFeB at the interface (see lower panels of Figure~\ref{Fig1}c for a zoom-in), though we observe CoFeB crystallites in the bulk of the film (not shown). In sharp contrast, we observe almost comprehensive crystallization of CoFeB in the case of the 60~\AA~film, especially at the barrier-ferromagnet interface. Together, the XRD and HRTEM data strongly advocate that thicker films ($d~\gtrsim$~500~\AA) do not crystallize completely after the anneal, especially at the interface with amorphous AlO$_\textrm{x}$, and consequently show a TSP similar to that of as-deposited amorphous CoFeB. On the contrary, thinner films crystallize virtually completely, and the TSP of crystalline CoFeB at its interface with AlO$_\textrm{x}$ manifests its intrinsic value. Note that the interface sensitivity of the TSP~\cite{leclair01b} is implicitly demonstrated within this inference. Furthermore, consistent with the observations of Takeuchi \textit{et al.}~\cite{TakeuchiJJAP}, in crystalline films, the out-of-plane grain size is limited by the film thickness, while the in-plane grain size (150-200~\AA) is similar to that observed in thicker films. As anticipated for such a Co rich composition, high angle XRD and Fourier transform of HRTEM images also confirm that CoFeB crystallizes in a highly (111) textured fcc~structure.

\begin{figure}[!t]
    \vspace{-2mm}\begin{center}
      \includegraphics[width=8.7 cm]{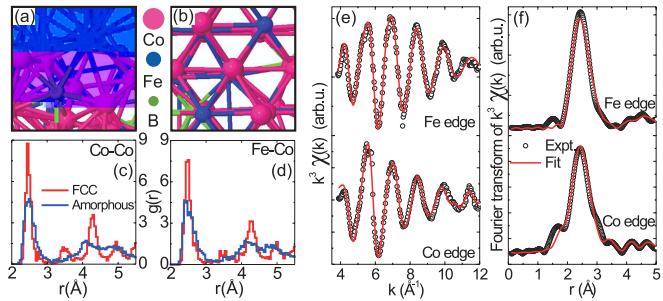}
      \vspace{-6mm}\caption{(color online)~(a) Representative amorphous and (b) fcc structures. Calculated pRDFs for Co-Co (c) and Fe-Co (d). Measured $k^3$-weighted EXAFS oscillations on Fe and Co K edges (e) and corresponding FT (f) for the amorphous films.}
      \label{Fig2}\vspace{-6mm}
    \end{center}
\end{figure}

Having established that the lowering of the CoFeB TSP is closely related to its crystallization, we embark on first-principles calculations using density functional theory within the generalized gradient approximation~\cite{gga}. The self-consistent electronic structure and interatomic forces are calculated with the projector augmented wave method~\cite{blo}~using the Vienna~\textit{ab-initio}~molecular dynamics package (VASP)~\cite{vasp1}. For reliable determination of the amorphous structure, the ensemble is heated above its melting point and equilibrated in the liquid state for time periods long enough to allow diffusion beyond one lattice spacing, and then rapidly quenched to form the amorphous state. Structural and electronic properties of two 108 atom ensembles are compared to three 54 atom ensembles for further verification and statistics. It is noteworthy that ensembles without B atoms do not quench in an amorphous structure, indicating the key role played by $\sim$7 at.~\% B in rendering CoFeB amorphous. In the fcc~case, the atoms are randomly placed in nominal positions of an fcc~lattice, and then allowed to relax. The total energy of the amorphous ensembles was invariably found to be higher than that of the distorted fcc~ensembles, consistent with the fact that as-deposited amorphous films crystallized after an anneal.

Representative structures of one amorphous and one fcc~ensemble are shown in Figure~\ref{Fig2}a~and~\ref{Fig2}b together with the partial radial distribution functions (pRDFs - Figure~\ref{Fig2}c and~\ref{Fig2}d). Irrespective of the size of the unit cell (108 or 54 atoms), the pRDFs show no significant difference in the inter- or intra-atomic coordination up to r~=~5.5~\AA, indicating that a 108 atom ensemble is of sufficient size. To gain insight in the atomic structure of amorphous films, EXAFS measurements were performed on Co and Fe K edges. The measured and fitted data are shown in Figure~\ref{Fig2}e and the corresponding Fourier transform (FT) in Figure~\ref{Fig2}f. The oscillations seen in Figure~\ref{Fig2}e are characteristic of disordered solids where usually the first coordination shell is the largest contributor to the fine structure, as is evident in the single peak dominating the FT. Keeping in mind the difficulties in fitting an amorphous structure, the fit to the oscillations is well within acceptable limits. More importantly, the fitted EXAFS data are in very good agreement with the coordination number and distance to the first and second shell that we find from the molecular dynamics. The fitted third coordination shells too agree fairly well with those obtained using molecular dynamics.

\begin{figure}[!t]
    \begin{center}
      \includegraphics[width=8.7 cm]{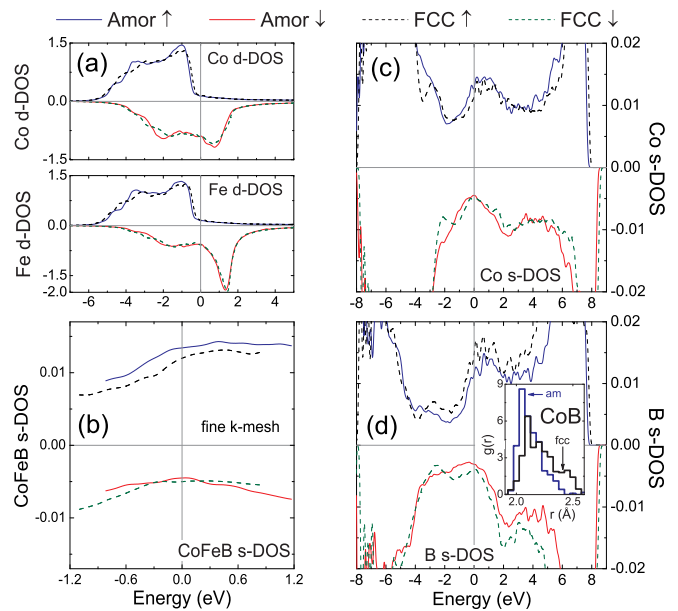}
      \vspace{-5 mm} \caption{(color online)~(a)~Element-specific d-DOS for Co and Fe.~(b)~total s-DOS on a fine~\textit{k}-mesh for CoFeB.~(c)~Co s-DOS and (d) B s-DOS. Inset in (d) shows pRDF for Co-B.}\vspace{-6 mm}
      \label{Fig3}
    \end{center}
\end{figure}

The calculated d-DOS for the amorphous and the fcc~alloy (see Figure~\ref{Fig3}a) show that both Fe and Co are in a strong ferromagnetic state with the majority channel completely filled. This is not surprising in the case of Fe considering the self-consistent density functional calculations of Schwarz~\textit{et al.}~\cite{SchwarzJPFMP14} on Co$_{100-\textrm{x}}$Fe$_\textrm{x}$, which show that the Fe magnetic moment increases with increasing number of Co nearest neighbors, and is largest when Fe has no Fe nearest neighbors. Comparing the d-DOS, both for Co and Fe, the d-band width is observed to be slightly lower in the amorphous case as compared to the fcc~case. This follows from the increase in the average Co-Co and Fe-Fe distance in the amorphous case (Figure~\ref{Fig2}c~and~\ref{Fig2}d) where the first coordination shell looses $\sim$1 atom and the second coordination shell around 3.5~\AA~is almost completely wiped out in comparison to the fcc~case.

Considering the amorphous nature of the barrier, one might argue that $k_\parallel$ conservation is highly unlikely in tunneling through AlO$_\textrm{x}$. In the first instance, if one neglects any issue related to the barrier or interface electronic structure, the spin polarization of s-like electrons, which have been experimentally shown~\cite{YuasaScience,YuasaPRLCr}~to dominate tunneling through AlO$_\textrm{x}$, is the only quantity which needs consideration. Table~\ref{Tab1} shows the calculated average s-electron SP at the Fermi level ($E_F$) for Co, Fe and B in the amorphous and fcc~case. Assuming that the concentration at the interface is similar to that in the bulk, we obtain the alloy SP by weighting these individual SPs with their concentrations~\cite{Meservey}. The last columns of Table~\ref{Tab1} compare the measured TSP to the calculated SP of the CoFeB alloy. For both the amorphous and fcc~case, the calculated SPs of 50$\pm$0.2\% and 41$\pm$0.5\% are in surprisingly good agreement with the measured TSPs of 53$\pm$0.5\% and 44$\pm$0.5\%, respectively. Most strikingly, the difference of $\sim$9\% between the two~\textit{measured}~TSP values is directly reflected in the calculations as well, indicating that this difference might arise from the disparity in the band structure of \textit{bulk} amorphous and fcc~CoFeB. It is noteworthy that in the case of the 5 amorphous and 2 fcc~unit cells studied, the values of the element-specific and the alloy SPs are remarkably similar from one unit cell to another. The errors in Table~\ref{Tab1} are deduced from the variations in the element-specific SPs under a coarse and a fine sampling of \textit{k}-space for the two 108 atom unit cells.

Interface bonding effects have been calculated to have pronounced effects on the TSP~\cite{TsymbalJAP97}. However, given the amorphous nature of AlO$_\textrm{x}$, these are rather difficult to predict, and in reality, they are an average over the configuration space at a disordered interface. We estimated the impact of the stronger bonding expected for B and Fe as compared to Co with oxygen at the interface, using an approach similar to Kaiser \textit{et al.}~\cite{KaiserPRL94}. Here too we did not see any significant deviation from the calculated SP values of Table~\ref{Tab1}. Given (1) the very good agreement between the SP of the bulk s-DOS with the measured TSP, (2) the striking agreement between the predicted and measured difference in the TSP of amorphous and fcc~CoFeB, and (3) the disordered structure of both the electrode and the barrier, one might wonder whether a better quantitative agreement can be achieved by going into further complexity.

\begin{center}
    \begin{table}[!t]
    \caption{Calculated s-SP and measured TSP values (in \%).}
        \begin{tabular*}{0.47\textwidth}{@{\extracolsep{\fill}}| p{14.45 mm}| c | c | c | c | c | c |}
        \hline
        \textbf{~~Struc.} & \textbf{Co} & \textbf{Fe} & \textbf{B} & \textbf{avg. SP} & \textbf{avg. SP} & \textbf{exp. TSP}\\
        & & & & \footnotesize{without B} & \footnotesize{with B} & \\
        \hline
        \textbf{\textit{a}-CoFeB} & 49.6 & 47.7 & 58.6 & 45.5 & 50.0$\pm$0.2 & 53$\pm$0.5 \\
        \hline
        \textbf{\textit{c}-CoFeB} & 40.5 & 39.9 & 54.5 & 37.4 & 41.4$\pm$0.5 & 44$\pm$0.5\\
        \hline
        \end{tabular*}
        \label{Tab1}
   \end{table}
\end{center}

\vspace{-10 mm}

Figure~\ref{Fig3}b shows the total s-DOS of amorphous and fcc~CoFeB, which confirms the higher SP of the amorphous alloy as given in Table~\ref{Tab1}. If one compares the element specific s-DOS for amorphous Co (and Fe - not shown) to fcc~Co (and Fe) in Figure~\ref{Fig3}c, the anti-bonding s-states of fcc~Co (and Fe) are pushed towards higher energy for both spin-channels. Increased s-d hybridization due to an increase in the first and second shell coordination of the fcc~alloy might be responsible for this (Figure~\ref{Fig2}c and~\ref{Fig2}d). Interestingly, the decrease in the s-electron SP of the fcc~alloy might be seen to primarily ensue from this spectral shift of the anti-bonding states towards higher energy, since $E_F$ lies on the slope of the increasing majority s-DOS, while lying in the deep minimum of the minority s-DOS. One notices from Figure~\ref{Fig3}b that the minority DOS also shows subtle changes, which provide a secondary contribution to the change in the s-electron SP. The impact of s-d hybridization can also be seen in the B s-DOS shown in Figure~\ref{Fig3}d. In our calculations we note that (1)~the B sp-states are highly spin polarized (s-SP$>50\%$; p-SP$>25\%$) as noted before~\cite{Coehoorn}, and (2)~the B sites attain a small negative magnetic moment ($\sim$~0.1$~\mu_{B}$) consistent with earlier work~\cite{HafnerPRB94}. This high polarization is a direct consequence of the hybridization of the B sp-states with the Co/Fe d-states forming covalent bonding states below $E_F$ and anti-bonding states above~\cite{Kanamori}. From the pRDFs of Co-B (see inset Figure~\ref{Fig3}d) and Fe-B (not shown), one notes that the peak in the first coordination shell around 2.1~\AA~is larger in the amorphous case as compared to the fcc~case. Consequently, for amorphous CoFeB, this leads to increased sp-d hybridization and the anti-bonding s-states of B are shifted to higher energy as seen in Figure~\ref{Fig3}d. Here, however, the spin polarization compared to the fcc~case increases due to the lower minority s-DOS at $E_F$. Moreover, we stress that the polarization of B s-states has a direct impact on the TSP. The fifth column in Table~\ref{Tab1} shows the calculated average SP of the alloy when the B atoms are considered unpolarized. The obvious disagreement with the measured TSP is an indication of the importance of highly spin-polarized B atoms at the interface.

In summary, we show that in AlO$_\textrm{x}$ based junctions, the TSP of amorphous CoFeB is larger than that of fcc CoFeB. Calculations of the atomic and electronic structure of amorphous and crystalline CoFeB yield s-electron SP values in remarkable agreement with experiment. These observations demonstrate that the electronic structure of the electrode has a marked impact on tunneling, and the electronic structure and SP of such a complex ternary amorphous/crystalline alloy can be genuinely calculated.

\textbf{Acknowledgements}~This research is supported by NanoNed, a Dutch nanotechnology program of the Ministry of Economic Affairs, by FOM, by STW via NWO-VICI grants, and by IP3 project of the European Commission: ESTEEM - Contract 0260019. We gladly acknowledge Reinder Coehoorn for fruitful discussions.

\end{document}